# PC 2 Phone Event Announcer

BEHRANG PARHIZKAR
Faculty of Information & Communication Technology, LIMKOKWING University
Cyberjaya, Selangor, Malaysia
hani.pk@limkokwing.edu.my

ABDULBASIT MOHAMMAD ABDULRAHMAN ALAZIZI
Faculty of Information & Communication Technology, LIMKOKWING University
Cyberjaya, Selangor, Malaysia
Basit5050@yahoo.com

MOHAMMAD NABIL SADEGH ALI
Faculty of Information & Communication Technology, LIMKOKWING University
Cyberjaya, Selangor, Malaysia
Nabil1420@yahoo.com

ANAND RAMACHANDRAN
Faculty of Information & Communication Technology, LIMKOKWING University
Cyberjaya, Selangor, Malaysia
anand@limkokwing.edu.my

SUJATA NAVARATNAM
Faculty of Information & Communication Technology, LIMKOKWING University
Cyberjaya, Selangor, Malaysia
sujata@leadership.edu.my

*Abstract*-Nowadays, mobile phones are indispensable devices; it has become a trend now that college and university students are owners of such devices and this particular factor plays a very important role behind the coming up with the proposed system. "PC 2 Phone Event Announcer", is the name of the new proposed system suggested to solve the existing communication problem between the College staff and students. As the name suggests, it can be deduced that the system will involve computers and phones, more specifically mobile phones.

I. INTRODUCTION

Mobile phones play an essential role in the mobile communication of the modern world. They make a person easily accessible. 'PC 2 Phone Event Announcer' makes it possible to access a group of persons in just a few clicks. The system will permit lecturers, records office staff and the librarian to contact students by sending SMS through their computers. The records office and the library will have a special feature called Autorun. This feature will do the job of the specified staff members by automatically scanning through the database for any late payments or overdue books and send SMS or emails to the concerned students.

In this paper we are going to conduct a study of the current system on the basic definition and concepts about Pc 2 Phone event announcer systems. Furthermore, we have done a small investigation about this topic. In the present scenario each and every person uses mobile phones. They have become a trend, students from every college and university are the owners of mobile phones and this particular factor plays a very vital role behind this project. Mobile phones play an important role in the mobile communication of the modern world. They make a person to contact with others very quickly.

II. RELATED WORK

In this ever increasing world of communication technology, the use of computers and cell-phones is getting wider and wider day by day. From the time the cell-phones were introduced to the market, the PCs and hand-phones have been increasingly sharing the similar applications and features. Nowadays, although being dependant from each other, they're being interconnected and inter-dependant in several ways. The following literature review tries to brief on several different studies that have been done on this interconnectivity of PCs and cell phones in different respects and also on different ways that cell-phone services have facilitated people's lives.

A research is conducted to find out how SMS technology can be used for as means of quick communication in particular the bulk SMS services used in organizations. We will go through many articles on SMS technology utilization by various organizations as means of communication.

EMERGENCY TEXT COMMUNICATIONS:

Being an absolutely useful communicational tool, text communications can specially be used in cases and places where there's no possibility of using voice calls. Wonsang Song, Jong Yul Kim, and Henning Schulzrinne, Piotr Boni and Michael Armstrong focused on the advantages of using text communications like IM and SMS in Emergency services. They focused on such features like being fast in service, text-based which can be accompanied with multimedia and Automatic Geographic Location in their article. Among the advantages of this service they pointed at its easy usability especially for the deaf and in places in which there's no chance of having a voice call. Also they mentioned that using this service, the current network can be maintained and it does away with the need to broaden the network bandwidth and so is cost-efficient [1].

AUTOMATED SECURED APPLIANCES:

Women University Rawalpindi of Pakistan has studied on a remote SMS-based system of controlling home appliances while the user is away from home in 2009. Using this system, users can have a global access to their home appliances and this feature enables them to automate their homes and secure them against intrusion with a relatively low cost. This is done through SMS using GSM technology. Since this system is wireless it's so adaptable and cost-efficient. Some technologies used here are GSM, RF (Radio Frequency), GPRS Modem, and Cell-phone via serial port RS 232 [2].

Also Fdhil T. Aula (2005) studied the same issue and discussed the techniques of using SMS for controlling home appliances through PC Parallel Port Interfacing [3].

SMS INTEGRATION IN ENTERPRISES:

In his article, Daniel Mavrakis, have reviewed different ways that PCs and cell-phones can be integrated into firms' routine chores using SMS. The featured mentioned by him are Alarm massages, Remote Control, Machine-To-Machine Interface, and Location Retrieval [4].

E-HEALTH ENHANCEMENT USING MOBILE COMMUNICATION:

Healthcare industry's increasing use of Mobile technology has been sparkling healthcare initiatives round the world. V. Dinusha and K.D Arunatileka focused on those aspects of mobile communication used in e-health like, sending scheduled SMS, replying and auto forwarding and RS232Serial Data cable. The technologies used here are Bluetooth, GSM, GPRS, EMR (Electronic Medical Record), infrared port, Serial cable connection, and RS232 serial data cable. Speaking about the advantages of using mobile communication in e-health, they believe that it enables, excels and aids accurate predictions, exact analysis, updates on diseases via SMS, timelier, a more comprehensive public health information and improved system of diagnosis disease tracking [5].

SMS-BASED PERSONAL ELECTROCARDIOGRAM MONITORING SYSTEM:

Ashraf A. Tahat studies the advantages of using SMS at the time when there's an emergency with Electrocardiogram Monitoring System. Features and technologies used in this article are Bluetooth Transceiver, The Microcontroller, Body Temperature Sensor, GPRS, EDGE, 3G, and WIMAX. In case of an emergency, when there's no immediate access to a cardiologist, ECG electrodes can be attached to the patient's body and the necessary information will be sent to the cardiologist's hand-phone, PDA or PC and a fast diagnosis and decision can be made by him/her [6].

SMS SERVICE FOR SCHOOLS/CONFERENCES:

James Kadirire explains in his article the ways that teacher-student relationship can be facilitated through the use of SMS, sent from PC to a phone and visa a versa. Featured used in this article are Internet, Email, Auto alerts, and Wierless and the technologies are JavaServlets, SMS Gateway, Unix Platform, and GSM Modem. The advantages of using such systems explained by Dr. Kadirire the facilitation of Presentations and Conference, the ability to maintain Instant Services, and elimination of cable usage due to the use of wireless systems [7].

SMS-BASED E-GOVERNMENT MODEL:

E-Governance has been getting a growing concern for the governments round the world especially in the developed and developing countries. It can be of great help to governments because of its potential for boosting up the official processes and doing away with much of bureaucracies. Toney Dwi and Robert proposed in their article some current technologies for the implementation of SMS-based systems in the application e-governance model. Some features used in their SMS-Based E-Government Model are Event-Based and Scheduled-Based SMS and the used technologies are SMS Cell Broadcast (CBS), SMSC and HTTP Protocols, ISP or ADSL, and Short Message Driver. Some of the advantages of this model mentioned by Susanto and Goodwin are the receipt of alerts at scheduled times (for scheduled

SMS), user's ability to request services by SMS, and the probability of having a Personal Profile [8]. This model maintains that local authorities can enhance their SMS-based governance through five levels: Notification, Presentation, Communication, Transaction, and Integration. Each level is a kind of e-government service that can be offered using SMS services.

MOBILE PHONES IN EDUCATION:

Michael Yerushalmy & Oshrat Benzaken has reviewed the ways that Mobile phone services can facilitate the education. Technologies used here are GPRS Modem, 3G, WiFi, J2ME, Medlets, and IMode. These technologies are used to send Instant Messages to mobile phone, and using those technologies this sending of IMs can be done with a PC using a GPRS Modem. The main advantage of using this technology is coming to the users' present location and their availability [9]. Classmates can easily get in touch with their peers and teachers, do a group discussion, ask for personalized information and teachers can check the progress of their students with no need of physical presence.

Saranphong Pramsane & Ridwan Sanjaya also believes that a PC to phone communication through SMS can be used in educational systems. The use of this system would result in closer relationships between teachers and students. They use technologies like WAP, GPRS and GSM Modem, HTTP, SMPP. The advantages of using such PC to phone messaging technologies are remote access, reliability in packet transfers, and the fact that HTTPs are acceptable in all browsers. Also they've considered the fact that not only texts but also multimedia files can be transferred from PCs to hand-phones using such technologies. This issue is of high importance and value in educational systems. [10] Tutors can send students whatever materials they need and this way of transmittance of the data is so cost-effective and efficient which any student can afford it.

Yet other researchers studied other different applications of PC-to-phone interconnections and mobile phone services. O. Awodele, E. R. Adagunodo, A.T. Akinwale, S. Idowu and M. Agbaje (2009) explained the use of cell-phone for delivering the exam results through SMS [11].

Didier Chincholle, Michael Bjorn, Christian Norlin & Morgan Lindqvist (2006) researched about the development of a user-centered IMS-based system through a user-oriented approach [12]. Rajive Chakravorty spoke about the limitations of using GSM and GPRS technologies in PC-to-phone communications and the advantages of using a TCP (Transmission Control Protocol) over GPRS [13]. In a similar study, Laurie Butgereit explained the usefulness of GPRS technology in Instant Messaging and how IMs can be sent using GPRS. [14] Neseem Alrawi worked on the use of use of Mobile SMS through GPRS and Java technologies in computer event communications [15]. Kjell suggested the use of Bluetooth and J2ME in sending and receipt of SMS and IM from PC to phone [16]. Also Klinkman studied the use of GSM Control SMS Gateway in sending message to hand-phones from PCs. [17] Finally, In an interesting study, Mahesh Gogineni & Aishwarya Lackshmi discussed the techniques and advantages of SMS in mobile phone that are capable being applied to urban microfinance [18].

### III. PROPOSED SYSTEM

The proposed system will work easily; the lecturers will have a system that will **enable** them to send SMS to their students from the web. They will have a Login ID and a password allowing them to access to the system. A student database consisting of student information such as their phone numbers, ID and name will be provided in order to facilitate the job of the lecturer to choose the students. After lecturer successfully login he can send Pc 2 Phone Event Announcer information to students by SMS. They will also have access to their respective timetables that will help for student retrieval as well (Fig. 1).

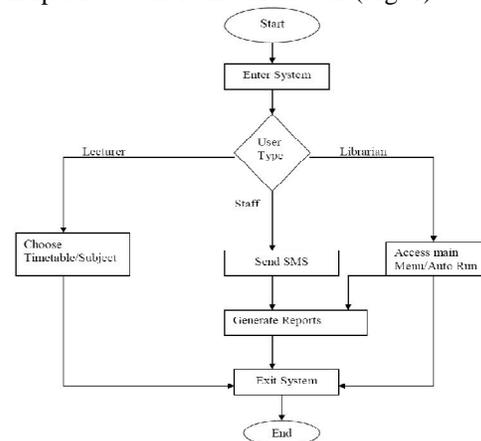

Figure 1: Proposed System Algorithm

There will be another sub-system in the library and in the records office. This system will have two options. The system in the records office will give the staff two options. One is to let the staff sort out which students have not paid their tuition fees yet and send reminders manually. The other one is the Autorun function. This function is about the system performing that task. The system will go through the student database and identify the students who have not settled their tuition fees yet. The system will then prompt the user for approval before sending out pre-formatted notification emails or SMS to the sorted students (Fig. 2).

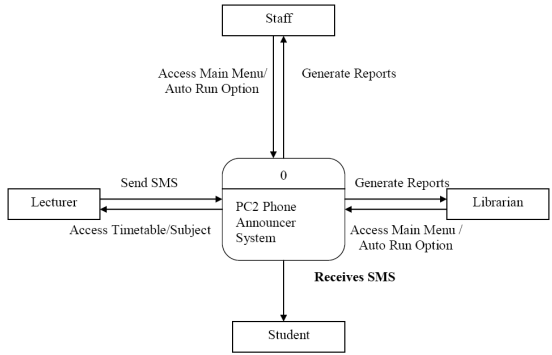

Figure 2: Proposed System Sub algorithm

A similar system will run in the library but the difference is that it will scan the database for overdue books and fines (Fig. 3).

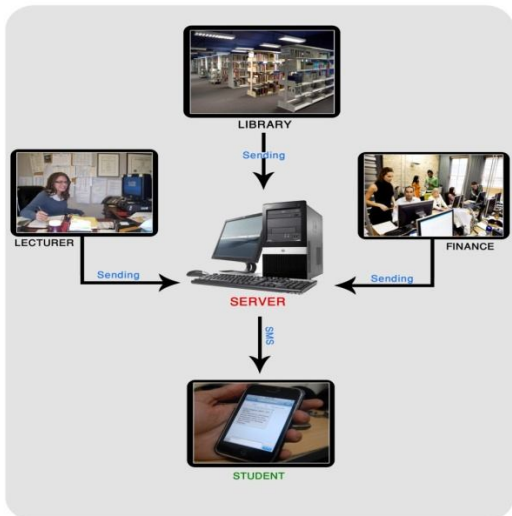

Figure 3: Proposed System Schema

### CONCLUSION

Going through many articles on SMS technology as a means of communication to send bulk SMS, has helped in gaining knowledge about SMS technology usage in various organizations. We have gained knowledge about how SMS technology can be used as a means of communication between students, lecturers and management. Apart from SMS service being just as a means of communication how best the information available with individuals and organizations can be communicated to students in universities is conducted with a view to pass valuable information that benefits the students.

After analyzing the system and collecting the feedback from the lecturers, students on the existing system, we have come to this conclusion that PC 2 Phone Event Announcer is the best solution to overcome the existing communication problems in the Universities.

### FUTURE WORK

Although the application of mobile phone technologies is not something new in the academic environments, but the provision of college (Lecturers, Records office and Library Staff) related services is still considered new.

More research should be done in this area where the concentration is put on the amount of information that needs to be filtered and delivered in this new environment. The challenge in research is the type of information services that the industries should deliver and find solutions to the versatile limitations existing in mobile phone technology in order to ensure satisfactory services to the users.

Firms need to work out the best ways to serve the users by the use this new mode of communication. They need to have constant collaboration with the researchers in the ICT fields, and also with the telecommunication providers to run a good study that looks into the appropriateness of industry functions that are to be delivered in the smaller interface environment.

The consequences of such study should be used in the actual execution of the services. In addition, the research conducted should also investigate the more advanced reference and information retrieval services and online public access with the help of the wireless technology.

### ACKNOWLEDGMENT

We would like to express our sincere gratitude to Arash Habibi Lashkari, PHD candidate of UTM University for his supervision and guidance. Also, we would like to express our appreciation to our parents and all the teachers and lecturers who help us to understand the importance of knowledge and show us the best way to gain it.